\documentclass[usegraphicx,usedcolumn,usenatbib]{mn2e}
\usepackage{xspace}
\usepackage{color,epsfig,rotating,amssymb,longtable}
\usepackage{array,colortbl}
\usepackage[colorlinks=true,linkcolor=magenta,citecolor=blue,breaklinks=true]{hyperref}

\usepackage{ifpdf}

\ifpdf
\usepackage{epstopdf}
\else
\usepackage[dvips]{graphicx}
\usepackage{lscape}
\fi

\def\kms{\,km\,s$^{-1}$\xspace}

\def\HI{H{\sc i}\xspace}
\def\HII{H{\sc ii}\xspace}

\def\OIII{O{\sc iii}}

\def\Msol{$M_\odot$\xspace}

\usepackage{color}

\definecolor{dgreen}{rgb}{0,.5,.1} 
\definecolor{pink}{rgb}{.9,.2,.5}  
\definecolor{orange}{rgb}{.9,.4,0} 
\definecolor{darkred}{rgb}{.545,0.0,.0}

\begin{document}

\title[Kinematical structure of CNSFRs]
{Implications of the kinematical structure of circumnuclear star-forming regions on their derived properties} 
\author[G. F.\ H\"agele et al.]
{Guillermo F.\ H\"agele$^{1,2,3}$\thanks{e-mail: ghagele@fcaglp.unlp.edu.ar}, 
\'Angeles I.\ D\'{\i}az$^{3}$, Roberto Terlevich$^{4,5}$,  
Elena Terlevich$^{4}$\thanks{Visiting astronomer at IoA, University of
  Cambridge, UK},
\newauthor 
Guillermo L.\ Bosch$^{2}$ and
M\'onica V.\ Cardaci$^{1,2,3}$
\\
$^{1}$ IALP-Conicet, Paseo del Bosque s/n, 1900 La Plata, Argentina.\\ 
$^{2}$ Facultad de Ciencias Astron\'omicas y Geof\'{\i}sicas, Universidad Nacional  
de La Plata, Paseo del Bosque s/n, 1900 La Plata, Argentina.\\ 
$^{3}$ Departamento de F\'{\i}sica Te\'orica, C-XI, Universidad Aut\'onoma de
Madrid, 28049 Madrid, Spain.\\
$^{4}$ Instituto Nacional de Astrof\'\i sica \'Optica y Electr\'onica, L.E.~Erro No. 1, Santa Mar\'\i a Tonantzintla,  Puebla, M\'exico\\
$^{5}$  Institute of Astronomy,University of Cambridge, Madingley Road, Cambidge CB3 0HA, UK\\} 

\maketitle


\begin{abstract}

We review the results of high dispersion spectroscopy of  17 circumnuclear star-forming regions (CNSFRs) in 3 nearby early spiral galaxies, NGC\,2903, NGC\,3310 and NGC\,3351. 

We find that single Gaussian fitting to the  H$\beta$ and [\OIII]$\lambda$5007\,\AA\ line profiles results in velocity dispersions 
around 32\,km\,s$^{-1}$ and 52\,km\,s$^{-1}$, respectively, while the IR
Ca{\sc ii} triplet cross-correlation technique provides stellar velocity dispersion  values close to 50\,km\,s$^{-1}$. 

Even though multiple kinematical components are present, the relation between gas velocity dispersion and Balmer emission line luminosity 
(L-$\sigma$ relation)    reproduces the correlation for disk giant \HII regions albeit with a larger scatter. The scatter 
in the L-$\sigma$ relation is considerably reduced when theoretical evolutionary corrections are applied suggesting that an age range is present in the sample of CNSFRs.

To analyse the observed complex profiles, we performed multiple Gaussian component fits to the  H$\beta$ and [\OIII]$\lambda$5007\,\AA\ lines obtaining optimal fits with two Gaussians  of different width. These best fits indicate that the narrower component has average velocity dispersion close to 23\,km\,s$^{-1}$  while the broader component  shows average values in the range 50-60\,km\,s$^{-1}$ for both lines, close to the observed stellar velocity dispersions. 
The fluxes of the broad and narrow H$\beta$ components are similar.
This is not the case for   [O{\sc iii}]$\lambda$5007\,\AA\ for which the broad components have higher fluxes than the narrow ones, thus producing a clear segregation in  their [O{\sc iii}]/H$\beta$ ratios.
 
We suggest a possible  scenario for understanding  the behaviour of CNSFRs in the  L-$\sigma$ and $\sigma_{gas}$-$\sigma_*$ diagrams involving  an inner gaseous disk responsible for the  narrow component of the emission lines.
 
Our main conclusion is that  the presence of different kinematical components with similar total fluxes in the emission line spectrum of CNSFRs raises important doubts  regarding properties of the ionized gas derived from global line ratios obtained with low resolution spectroscopy in star-forming regions in the central regions of early type galaxies. 
Given the ubiquity of these star-forming systems, ionized gas analyses should be done preferably from high dispersion spectra with high spatial resolution.


\end{abstract}

\section{Introduction}

Star formation in a circumnuclear configuration is of  common occurrence in early spiral galaxies with or without an active nucleus. 
These structures are frequently arranged in a ring pattern with a diameter of about 1\,kpc \citep{Morgan58,Sersic+65,Sersic+67} and their associated star formation rate (SFR) is considerably higher than the average observed in galactic discs. In fact, in the ultraviolet (UV), the luminosity coming from their young massive stars can dominate the observed circumnuclear emission even in the presence of an active nucleus \citep{Gonzalez-Delgado+98,Colina+02}. 

The properties of circumnuclear star-forming regions (CNSFRs) are different enough from giant extragalactic \HII regions that they can be considered to constitute a distinctive mode of star formation in galaxies (Kennicutt 1998). This distinct nature was fully revealed with the opening of the mid and far-IR spectral ranges \citep[see for example][]{Rieke+72,Harper+73,Rieke+78,Telesco+80}. CNSFRs look more compact and show higher peak surface brightness than disc \HII regions \citep{Kennicutt+89} while their optical spectra show the weak oxygen forbidden lines characteristic of a high abundance \citep[][hereafter D07]{Diaz+07}. 

It has been suggested that the substructures (sizes of the order of 10\,pc or less) found in circumnuclear star forming complexes correspond to massive (M $\sim$ 10$^5$ \Msol)  young (age $\lesssim$ 1\,Gyr) clusters. If this is the case, they can be modelled as single stellar population  applying population synthesis techniques and their properties such as age, chemical evolution and star formation histories could be derived with only small uncertainties due to discreetness related scatter \citep[see e.g.][]{Cervino+02} .
Although there have been some recent works devoted to the study of CNSFRs most of them have concentrated in deriving the SFR and ages of these structures and little has been done with respect to their structure and internal kinematics. 

In the last few years we have obtained spectroscopic data of CNSFRs in nearby
galaxies with a resolution enough to allow the measurement of stellar and
gaseous velocity dispersions that, combined with their sizes obtained from the
Hubble Space Telescope (HST)
imaging, have yielded values for the dynamical masses of these objects
\citep[][hereafter H07, H09 and H10, respectively]{Hagele+07,Hagele+09,Hagele+10a}. 
In this paper we present the results obtained from the joint analysis of the data and address several aspects related to the properties of these structures. In particular, we analyse the relations between stellar and gaseous velocity dispersions and the behaviour of the velocity dispersion-luminosity ($\sigma$-L) and mass-luminosity (M-L) relations for CNSFRs. We also draw attention to several concerns related to conclusions about gas properties of central regions of galaxies, like average chemical abundances or characteristics of ionizing sources, derived from data with low spatial and spectral resolution.


\section{Summary of data and derived parameters}
\label{obs-sample}


The CNSFRs studied in the present work correspond to three early type barred spiral galaxies, NGC\,2903, NGC\,3310 and NGC\,3351 whose properties are given in Table \ref{propgal}. 

%

\begin{table}
\begin{center}
\caption{The galaxy sample.}
 \begin{tabular}{@{}lccc@{}}
\hline
Property                     & NGC\,2903    & NGC\,3310    & NGC\,3351     \\
\hline				             
RA (2000)$^a$                &  09 32 10.1  & 10 38 45.9   & 10 43 57.7     \\
Dec.\ (2000)$^a$             & +21 30 03    & +53 30 12    & +11 42 14      \\ 
Morph.\ type                 & SBbc         &  SABbc       & SBb            \\
Distance (Mpc)               &  8.6\,$\pm$\,0.5$^b$     &  15\,$\pm$\,1$^a$       &    10.05\,$\pm$\,0.88$^c$   \\
pc/arcsec                    &  42\,$\pm$\,2          &  73\,$\pm$\,5          &    49\,$\pm$\,4          \\
B$_{T}$ (mag)$^a$             &  9.7\,$\pm$\,0.1        &  11.2\,$\pm$\,0.1        &    10.5\,$\pm$\,0.1       \\
E(B-V)$_{gal}$(mag)$^a$        & 0.031\,$\pm$\,0.01        &  0.030\,$\pm$\,0.01       & 0.028\,$\pm$\,0.01          \\
\hline
\multicolumn{3}{l}{$^a$~\cite{deVaucouleurs+91}}\\
\multicolumn{3}{l}{$^b$~\cite{Bottinelli+84}}\\
\multicolumn{3}{l}{$^c$~\cite{Graham+97}}\\
\end{tabular}
\label{propgal}
\end{center}
\end{table}


For 17 star forming complexes we obtained high resolution spectroscopy in the
blue (4779 to 5199\,\AA ) and red (8363 to 8763\,\AA ) with resolutions of
0.21 and 0.39 \AA\ per pixel, respectively, providing a comparable velocity
resolution of about 13\,km\,s$ ^{-1}$. The blue spectral region comprises the
emission lines of H$\beta$\,$\lambda$\,4861\,\AA\ and [O{\sc
  iii}]\,$\lambda$\,5007\,\AA\  and have been used to measure the gas velocity
dispersion. The red spectral region comprises the stellar absorption lines of
the Ca{\sc ii} triplet (CaT) at $\lambda\lambda$\,8494, 8542, 8662\,\AA, and
have been used to measure the stellar velocity dispersion. The two spectra
have been obtained simultaneously using the double arm
Intermediate-dispersion Spectrograph and Imaging System (ISIS) attached to the
William Herschel Telescope and therefore we are confident that they refer to
the same regions. More details about the observations can be found in our
previous works (H07, H09, H10).


The velocity dispersions of the gas, {$\sigma_{gas}$}, were calculated as
\begin{equation}
\sigma_{gas}\,=\,\sqrt{\sigma_m^2\,-\,\sigma_i^2}
\end{equation}
\noindent where $\sigma_m$ and $\sigma_i$ are the measured and 
instrumental dispersions, respectively, {in km\,s$^{-1}$}. $\sigma_i$ was  measured directly from
the sky emission lines and is about 10.4\,km\,s$^{-1}$ at $\lambda$ 4861 \AA . No thermal broadening correction was applied, thus our sigmas slightly overestimate the gas intrinsic dispersion.

\begin{figure*}
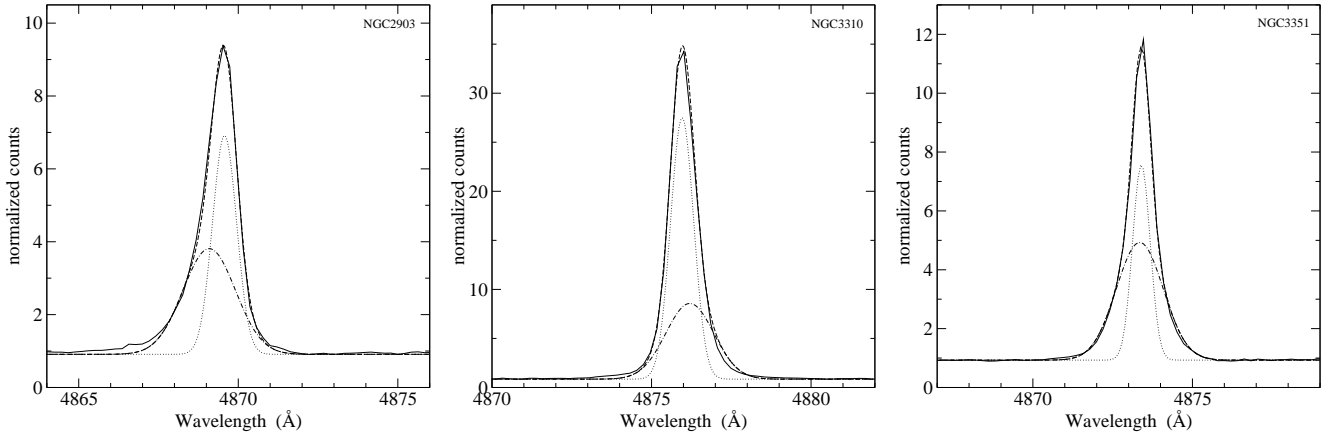

\centering
\includegraphics[width=.32\textwidth,angle=0]{figures/ngauss-Hb-R1-2903.eps}\hspace*{0.15cm}
\includegraphics[width=.32\textwidth,angle=0]{figures/ngauss-Hb-R4+R5-3310.eps}\hspace*{0.15cm}
\includegraphics[width=.32\textwidth,angle=0]{figures/ngauss-Hb-R2-3351.eps}

\caption[]{Optimal Gaussian fits to the H$\beta$ line for observed regions in the galaxies: NGC~2903, NGC~3351 y NGC~3310. The two different components labelled ``broad'' and ``narrow'' are shown.}
\label{ngauss}
\end{figure*}

The single component Gaussian fits performed to the H$\beta$ line  were not totally satisfactory. In fact,  in all cases, the optimal fit was found for two different Gaussian components. In most cases, similar components also provided optimal fits to the [\OIII] lines. 
Fig.\ \ref{ngauss} shows fits to the H$\beta$ lines in CNSFRs of each of the three galaxies studied as illustrative examples.


\begin{figure}
\centering
\includegraphics[width=.32\textwidth,angle=0]{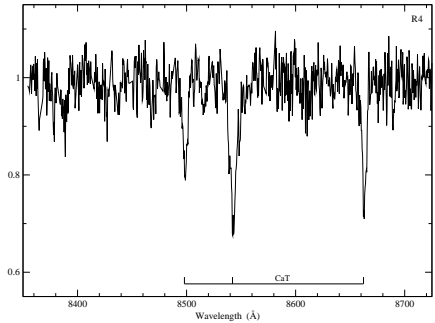}

\caption[]{Spectrum of region R4 in NGC~2903 in the IR CaII triplet spectral region. } 
\label{CaT}
\end{figure}

The stellar velocity dispersions, {$\sigma_*$,} were estimated from the CaT lines by cross-correlating the spectrum corresponding to each of the regions with stellar templates. An example of the CaT spectral region can be seen in Fig. \ref{CaT}.


When observed with the HST resolution, most of the observed  CNSFRs seem to be
a composite of more than one cluster with radii between 1.5 and 6.2\,pc. The
virial theorem has been applied to estimate upper limits to the dynamical
masses inside the half light radius for each identified knot under the
assumption that the clusters are gravitationally bound,  spherically symmetric
and have isotropic velocity distributions \textbf{ [$\sigma^2$(total)\,=\,3 $\sigma_{\ast}^2$]. 
The general expression for the virial mass of a cluster is 
\begin{equation}
M_{\ast}\,=\,\eta\,\sigma_{\ast}^2\,R/G
\end{equation}
where R is the effective gravitational
radius and $\eta$ is a dimensionless number that takes into account departures 
from isotropy in the velocity distribution and the spatial mass distribution, binary
fraction, mean surface density, etc.\ \citep{Boily+05,Fleck+06}. Following
\cite{Ho+96a,Ho+96b}, and for consistence with
H07, H09, H10 and \cite{HagelePhD}, we obtain the dynamical masses inside the
half-light radius 
using $\eta$\,=\,3 and adopting the half-light radius as a reasonable
approximation of the effective radius.}
Since our measured stellar velocity dispersion encompass each whole region to which the single clusters belong, their masses can be overestimated to some extent. However we think the mass overestimate cannot be large, because the equivalent widths of the CaT lines point to the spectral energy distribution of each region being dominated by the young stellar population belonging to the clusters rather than by the galaxy bulge. The adopted mass for each region has been obtained by summing up the individual masses of the single clusters. More details about these procedures can be found in  H07.

Table  \ref{disp} lists the velocity dispersions  in each CNSFR  together with the derived values of the dynamical masses. 
Column 1 gives the name of the region as coded in the original publication;
column 2 gives the stellar velocity dispersion ($\sigma_{\ast}$); column 3
gives the value of the gas velocity dispersion derived using a single Gaussian
fit ($\sigma_{s}$) to the H$\beta$ line; columns 4 and 5 give the
corresponding values derived by the two-component optimal fit that have been
labelled  narrow ($\sigma_{n}$) and broad ($\sigma_{b}$),
respectively. Columns 6 to 8 provide the same information for the [\OIII]
emission lines. Finally, column 9 gives the derived dynamical mass, {$M_{ast}$}. 

There are three objects with two sets of values listed in the table which correspond to different observations acquired in two different slit position angles.


\begin{table*}
\centering
{\footnotesize
\caption{Velocity dispersions and dynamical masses for the observed CNSFR}
\begin{center}
\begin{tabular} {@{}l c c c c c c c c@{}}
\hline
\hline
            &                     &               \multicolumn{3}{c}{ H$\beta$} & \multicolumn{3}{c}{[\OIII]}  \\
   
 Region & $\sigma_{\ast}$  & $\sigma_{s}$ &  $\sigma_{n}$  & $\sigma_{b}$  & $\sigma_{s}$ &  $\sigma_{n}$  & $\sigma_{b}$  & $M_{\ast}$  \\
\hline

\multicolumn{9}{c}{NGC\,2903}\\[2pt]

R1+R2 & 60$\pm$3 & 34$\pm$2 & 23$\pm$2 & 51$\pm$3  &  73 $\pm$ 8   & 26 $\pm$ 8 & 93 $\pm$  9  & 1816$\pm$48  \\
R1+R2 & 64$\pm$3 & 35$\pm$2 & 27$\pm$2 & 53$\pm$4  &  71 $\pm$ 9   & 27 $\pm$ 7 & 95 $\pm$ 10 & 2054$\pm$45  \\
R4       & 44$\pm$3 & 32$\pm$2 & 20$\pm$2 & 47$\pm$4  &  76  $\pm$ 10 & 35 $\pm$ 9 & 89 $\pm$  8  & 853$\pm$28   \\
R7       & 37$\pm$3 & 32$\pm$4 & 29$\pm$5 & 34$\pm$8  &  59  $\pm$ 10 & 17 $\pm$ 5 & 67 $\pm$  8  & 642$\pm$26  \\[2pt]
                                                                                                
\multicolumn{9}{c}{NGC\,3310}\\[2pt]                                                           
                                                                     	                   
R1+R2 &   80:           & 33$\pm$4 & 24$\pm$5 & 54$\pm$7 &  31 $\pm$ 4 & 22 $\pm$ 5 & 50 $\pm$ 4 &  911:                \\
R4        & 36$\pm$3 & 34$\pm$3 & 28$\pm$4 & 55$\pm$5 &  32 $\pm$ 3 & 26 $\pm$ 3 & 52 $\pm$ 4 &  886 $\pm$30   \\
R4+R5 & 38$\pm$3 & 27$\pm$4 & 22$\pm$4 & 46$\pm$5 &   22 $\pm$ 3 & 18 $\pm$ 3 &40 $\pm$ 2 & 1317 $\pm$41   \\
R6       & 35$\pm$5 & 30$\pm$3 & 23$\pm$3 & 54$\pm$7 &   28 $\pm$ 3 & 21 $\pm$ 3 & 56 $\pm$ 3 &  397 $\pm$33   \\
S6       & 31$\pm$4 & 27$\pm$3 & 20$\pm$3 & 47$\pm$4 &   26 $\pm$ 3 & 19 $\pm$ 3 & 47 $\pm$ 3 &  210 $\pm$17   \\
R7       & 44$\pm$5 & 21$\pm$5 & 18$\pm$4 & 41$\pm$4 &   17 $\pm$ 4 & 14 $\pm$ 3 & 36 $\pm$ 3 & 1413$\pm$60   \\
R10     & 39$\pm$3 & 38$\pm$3 & 26$\pm$2 & 54$\pm$2 &   40 $\pm$ 3 & 26 $\pm$ 3 & 59 $\pm$ 4 &  588 $\pm$28   \\[2pt]
J         &    ---         & 34$\pm$2 & 25$\pm$2 & 61$\pm$3 &     30 $\pm$ 2 & 22 $\pm$ 2 & 57 $\pm$ 3 &  ---                  \\

\multicolumn{9}{c}{NGC\,3351}\\[2pt]                                	                   
                                                                     	                   
R2    & 50$\pm$1 & 26$\pm$1 & 17$\pm$3 & 45$\pm$3 &  72 $\pm$ 7 & 21 $\pm$ 4 & 74 $\pm$ 5 & 129$\pm$6   \\
R2    & 51$\pm$6 & 29$\pm$3 & 16$\pm$2 & 43$\pm$2 &  69 $\pm$ 9 & 23 $\pm$ 5 & 76 $\pm$ 8 & 131$\pm$23  \\
R3    & 55$\pm$5 & 35$\pm$1 & 25$\pm$3 & 59$\pm$4 &  67 $\pm$ 7 & 28 $\pm$ 4 & 71 $\pm$ 4 & 417$\pm$31  \\
R3    & 59$\pm$7 & 39$\pm$5 & 24$\pm$3 & 59$\pm$3 &  70 $\pm$ 7 & 24 $\pm$ 6 & 74 $\pm$ 9 & 477$\pm$47  \\
R4    & 66$\pm$4 & 37$\pm$4 & 29$\pm$3 & 65$\pm$4 &  76 $\pm$ 8 & ---              & ---              &   87$\pm$12  \\
R5    & 47$\pm$4 & 34$\pm$2 & 30:            & 76:             &   56 $\pm$ 7 & ---             & ---              &   49$\pm$8   \\
R6    & 39$\pm$6 & 29$\pm$6 & 16$\pm$3 & 46$\pm$4 &   46 $\pm$ 7 & ---             & ---              &   434$\pm$39  \\[2pt]

\hline
\multicolumn{9}{@{}l}{Note: velocity dispersions in km\,s$^{-1}$, masses in
  10$^5$\,$M_\odot$. {The colons instead of the}}\\
\multicolumn{9}{@{}l}{{corresponding errors indicate
  that those quantities have large associated}}\\ 
\multicolumn{9}{@{}l}{{errors (greater than 40\%).}} 
\end{tabular}
\end{center}
\label{disp}
}
\end{table*}



For every region we have derived the mass of the ionising stellar population, ($M_{ion}$), from published values of the H$\alpha$ luminosity \citep{Planesas+97,Diaz+00a,Pastoriza+93} corrected for reddening assuming the galactic extinction law of \cite{Miller+72}
with R$_{\textrm v}$\,=\,3.2 and, when necessary, correcting for recent
changes in their distances. The solar metallicity {[12+log(O/H)\,=\,8.919]} single burst models by
\cite{Garcia-Vargas+95} were used. These models assume a Salpeter initial mass function \citep[][IMF]{Salpeter55}
with lower and upper mass limits of 0.8 and 120 $M_\odot$ and provide the number of ionising Lyman continuum photons [$Q(H_0)$] per unit mass of the ionising population, [$Q(H_0)/M_{ion}$].  This number decreases with the time evolution of the population, and is related to the computed equivalent width of the H$\beta$ line as \citep[e.g.\ ][]{Diaz+00c}: 

\begin{equation}
log \left[ Q(H_0)/M_{ion} \right] = 44.48 + 0.86 \log \left[ EW({\rm H\beta})\right] 
\end{equation}

The total number of ionising photons for a given region has been derived from the H$\alpha$
luminosities \citep{1995ApJS...96....9L}:

\begin{equation}
Q(H_0)\,=\,7.35\,\times\,10^{11}\,L({\rm H\alpha})
\end{equation}

The final expression for the derivation of $M_{ion}$ is then: 

\begin{equation}
M_{ion}\,=\,\frac{7.35\,\times\,10^{11}\,L({\rm H\alpha)}}{10^{44.48\,+\,0.86\,\log\left[EW({\rm H\beta})\right]}}
\end{equation}

The equivalent widths of the H$\beta$ line have been taken from \cite{Pastoriza+93} and D07. 
Given that the H$\beta$  equivalent widths are probably affected by an underlying older stellar continuum not belonging to the ionizing cluster itself, the listed masses for these clusters should be considered upper limits. On the other hand, no photon escape or absorption of  ionizing photons by dust has been considered, which would place lower limits to the mass. These two effects probably cancel each other to  a certain degree.
 
The mass of ionised gas ($M_{{\rm HII}}$) associated to each star-forming
region complex was derived from the H$\alpha$ luminosities, 
using the electron density (N$_e$) dependency relation given by
\cite{Macchetto+90} for an electron temperature of 10$^4$\,K.
\begin{equation}
M_{{\rm HII}}\,=\,3.32\,\times\,10^{-33}\,L({\rm H\alpha})\,N_e^{-1}
\end{equation}

The electron density for each region was taken from \cite{Pastoriza+93} and D07. For those regions of NGC\,3310 without electron density estimates a value of  N$_e$ equal to 100\,cm$^{-3}$, corresponding to the average value of other CNSFRs of this galaxy, was assumed.


The values of the H$\alpha$ luminosities for each CNSFR together with quantities derived from them  are given in Table \ref{lum}. 
Also in this table, and following the work by \cite{Relano+05},  we list the individual Gaussian component fluxes as fractional emission (EM$f$), in percentage,  relative to the total line flux. 

These have been calculated from the emission line fluxes estimated for each
H$\beta$ component from the parameters derived using the fitting from the {\sc
  ngauss} task of IRAF\footnote{Image Reduction and  
Analysis Facility, distributed by NOAO, operated by AURA, Inc., under
agreement with NSF.}. Individual component H$\alpha$ luminosities follow directly from those with the only assumptions that the reddening corrections are the same for both components (although this is not necessarily true, see \citealt{Hagele+12}) and that the EM$f$ remains constant across the whole nebula. It must be noticed that the slit is 1\,arcsec wide and the estimated diameter of the entire star-forming complexes from which the photometric fluxes were estimated, are about 2\,arcsec. 

As in the case of Table \ref{disp} there are several objects with two sets of values listed in Table \ref{lum} which correspond to two different observations as explained above.



\begin{table*}
\centering
{\footnotesize
\caption{H$\alpha$ luminosities and derived quantities, and H$\beta$
  equivalent widths.}
\begin{center}
\begin{tabular} {@{}l c c c c c c c c @{}}
\hline
\hline
 Region & $L({\rm H\alpha})_t$ &
 EM$f_n$ &  EM$f_b$ & $L({\rm H\alpha})_n$ & $L({\rm H\alpha})_b$ &
 $M_{ion}$ & $M_{{\rm HII}}$ &  EW(H$\beta$)\\
 \hline

\multicolumn{9}{c}{NGC\,2903}\\[2pt]

R1+R2  &  66.3$^a$ & 49 & 51 & 32.3 & 34.0 &  18.9 & 0.79     & 12.1  \\     
R1+R2  &     ---       & 57 & 43 & 38.1 & 28.2 &  ---  &  --- & 12.1  \\     
R4        &  38.9$^a$ & 32 & 68 & 12.6 & 26.3 &  24.6 & 0.48  &  4.8  \\     
R7        &  31.3$^a$ & 59 & 41 & 18.6 & 12.7 &  23.6 & 0.30  &  3.9  \\[2pt]

\multicolumn{9}{c}{NGC\,3310}\\[2pt]                                                           
                                                                     	                   
R1+R2   & 102$^b$  & 48 & 52 &  49.3 & 52.7 & 13.9 & 3.38 & 28.6         \\      
R4          & 144$^b$  & 58 & 42 &  83.5 & 60.5 & 17.6 & 4.78 & 32.4     \\      
R4+R5    & 218$^b$  & 62 & 38 &  136  & 82.0 & 21.4 & 7.24 & 41.7        \\      
R6          & 57.3$^b$ & 53 & 47 &  30.5 & 26.8 & 12.4 & 1.90 & 16.7     \\      
S6          & 62.5$^c$ & 54 & 46 &  34.0 & 28.5 & 17.4 & 2.07 & 12.5     \\      
R7          & 45.5$^b$ & 72 & 28 &  32.7 & 12.8 & 8.7  & 1.51 & 19.4     \\      
R10        & 45.5$^b$ & 39 & 61 &  18.0 & 27.5 & 15.7 & 1.51 &  9.7      \\[2pt] 
J             & 573$^b$  & 51 & 49 &  294  & 279  & 31.4 & 9.52 & 82.5   \\      

\multicolumn{9}{c}{NGC\,3351}\\[2pt]                                	                   
                                                                     	                   
R2    & 28.3$^a$ & 40 & 60 &  11.4 & 16.9 & 9.93 & 0.21 & 9.5        \\     
R2    & ---          & 35 & 65 & 10.0 & 18.3 & ---  & ---  & 9.5     \\     
R3    & 70.0$^a$ & 49 & 51 &  34.4 & 35.7 & 15.3 & 0.54 & 16.5       \\     
R3    &  ---         & 33 & 67 &  22.9 & 47.1 & ---  & ---  & 16.5   \\     
R4    & 23.2$^a$ & 48 & 52 &  11.1 & 12.1 & 6.23 & 0.25 & 13.0       \\     
R5    & 10.3$^a$ & 80 & 20 &   8.2  & 2.1  & 6.17 & 0.09 & 5.1       \\     
R6    &  7.5$^a$  & 46 & 54 & 3.4  & 4.0  & 8.88 & 0.07 & 2.3        \\[2pt]

\hline
\multicolumn{9}{@{}l}{Note: EM$f$ in percentage, luminosities in 10$^{38}$\,erg\,s$^{-1}$, 
  masses in 10$^5$\,$M_\odot$.}\\
\multicolumn{9}{@{}l}{$^a$From \cite{Planesas+97} corrected for the different
  adopted distances, and for }\\
\multicolumn{9}{@{}l}{reddening using E(B-V) from \cite{tesisdiego}.}\\
\multicolumn{9}{@{}l}{$^b$ From \cite{Diaz+00a}.}\\
\multicolumn{9}{@{}l}{$^c$ From \cite{Pastoriza+93}.}
\end{tabular}
\end{center}
\label{lum}
}
\end{table*}



\section{Discussion}
\label{discussion}

\subsection{Star and gas kinematics}

Given the emission line profile differences between permitted and forbidden lines already reported by us in previous works (H07, H09, H10), it  is important to study the relationship between  the velocity dispersion of the stars and the ionised gas for all the CNSFRs that have been observed at high dispersion and with similar instrumentation.

A quick inspection of Table \ref{disp} shows that while  the stellar velocity
dispersion in the CNSFRs ranges between $\sim $30 and $\sim $75\,\kms  with an
average value of 49\,\kms, the ionized gas velocity dispersion, as obtained
from the H$\beta$ emission line using a single Gaussian fit, shows a range
between $\sim $20 and $\sim $40\,\kms  with an average value of 32\,\kms.  On
the other hand, the [\OIII] velocity dispersion, also measured from a single
Gaussian fit, that ranges between $\sim $17 and $\sim $75\,\kms  with an
average value of 52\,\kms has values closer to the stellar velocity dispersion
perhaps indicating some extra source of broadening with respect to H$\beta$
(see a detailed discussion in H09). 

\begin{figure}
\centering
\vspace*{0.3cm}
\includegraphics[width=.45\textwidth,angle=0]{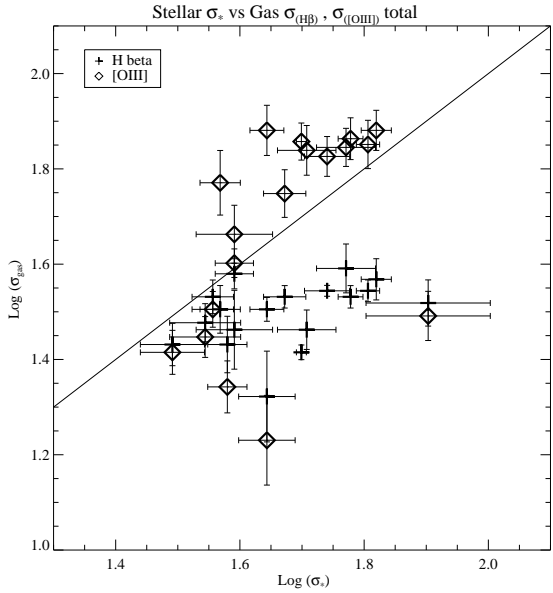}
\caption[]{Gaseous versus stellar velocity dispersions from single Gaussian fits. Pluses correspond to H$\beta$ while open diamonds correspond to [\OIII]. The continuous line represents:  $\sigma _{gas}$ $=$ $\sigma _{stars}$.}
\label{dispersions-comp-a}
\end{figure}

The systematic shift between the ionized gas and stellar velocity dispersion
is clearly seen in Fig.\ \ref{dispersions-comp-a} that shows the gas  velocity
dispersion measured by fitting a single Gaussian to the H$\beta$ and [\OIII]
lines versus the stellar velocity dispersion for all the regions.  Even though
the scatter is large 
it is  possible to see that while most of the CNSFRs with stellar velocity
dispersion below $\sim$40\,\kms (all but two belonging to NGC~3310) show
similar stellar, H$\beta$ and [\OIII] velocity dispersions, the CNSFRs with
stellar velocity dispersion above $\sim$40\,\kms show a clear dichotomy with
the [\OIII] profiles being systematically  broader than those of H$\beta$. 
\textbf{The Balmer emission lines are affected by the absorption produced by the
  underlying stellar population that depress the lines (see a detailed discussion about this effect
  in H07, H09 and H10). To minimise the errors introduced by the underlying
  population we defined a pseudo-continuum at the base of the line to fit
  the emission lines \cite[for details, see][]{Hagele+06}. Nevertheless, the presence of a conspicuous underlying
  population could still be affecting the $\sigma_{gas}$ derived from the H$\beta$
  emission lines. On the other hand, the fact that the regions with higher
  values of $\sigma_*$ have systematically larger disagreement with the
  $\sigma_{gas}$\,=\,$\sigma_*$ line could be related with that they are
  expected to be the most massive ones having the larger amount of stars and 
  the most conspicuous underlying stellar populations.}

 As mentioned above, most of the observed CNSFR emission line profiles show wings detectable in both the H$\beta$ hydrogen recombination line and the [\OIII] forbidden line. These wings can be attributed to the presence of a component  broader than the core of the line. To investigate the origin of this broad component we have assumed that it can be represented by a Gaussian profile. It should be kept in mind, however, that given that we do not have an a-priori scenario for the origin of the broad component, there is no  justification to assume a Gaussian profile either, apart from the fact that this assumption simplifies the analysis. 
 If we assume that the wings are due to stellar winds, the broad component profile should be that of an expanding shell, i.e. a very flat top profile in which case the fluxes of the broad component obtained with a broad Gaussian will be overestimated and the dispersion of the narrow component will be underestimated. Yet,  the best Gaussian fits involved two different components for the gas: a narrow component with velocity dispersions lower than the stellar one and a broad component with a velocity dispersion similar to or slightly larger than that measured for the stars. The resulting values of the multiple Gaussian fit to the observed profiles for both H$\beta$ and [\OIII]  narrow and broad components are listed in Table \ref{disp}  and Table \ref{lum}. 

\begin{figure}
\centering
\vspace*{0.3cm}
\includegraphics[width=.45\textwidth,angle=0]{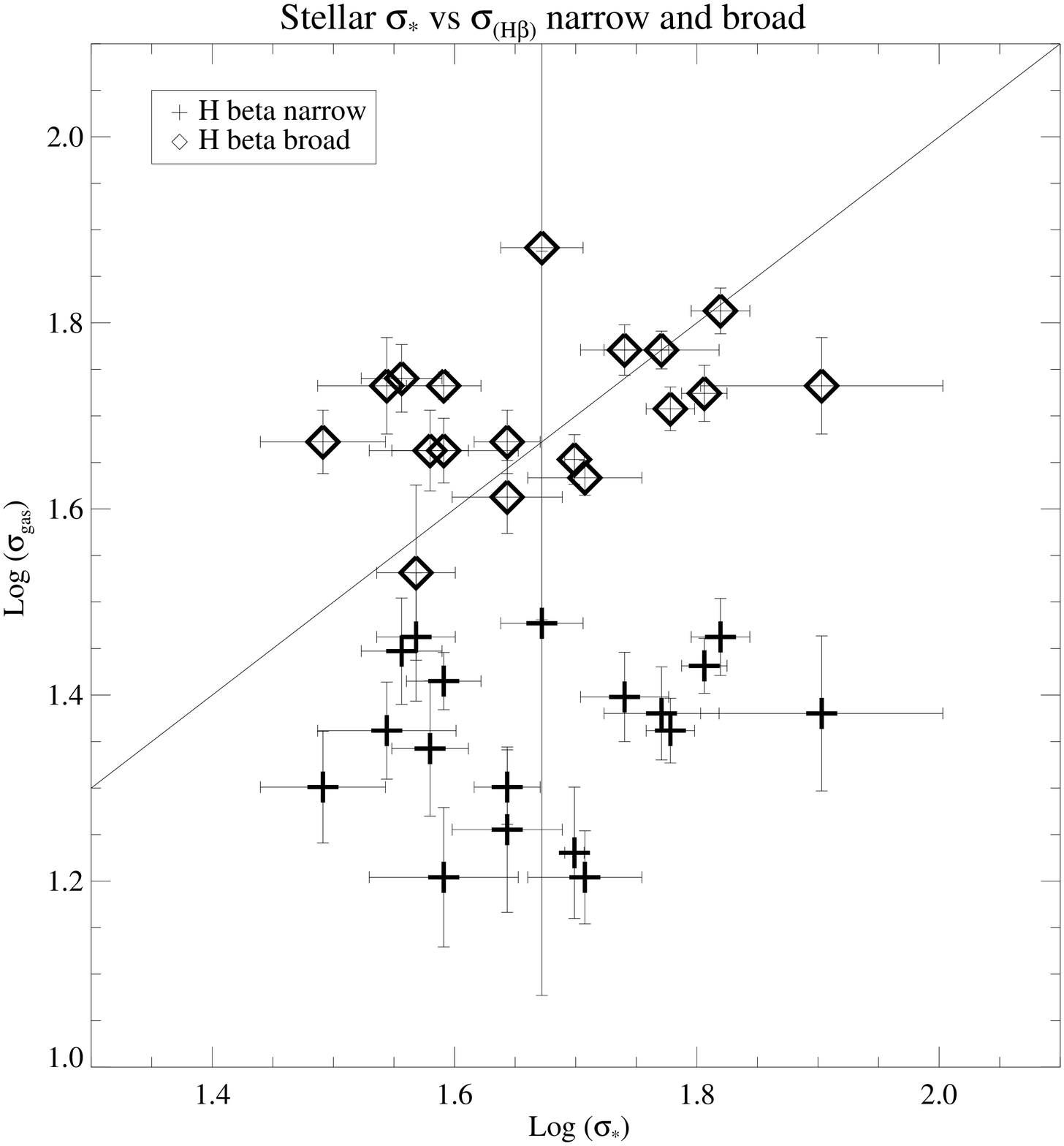}
\caption[]{Gaseous (H$\beta$)  versus stellar velocity dispersions for all the observed circumnuclear regions.  The continuous line represents:  $\sigma _{gas}$ $=$ $\sigma _{stars}$.}
\label{dispersions-comp-b}
\end{figure}

\begin{figure}
\centering
\vspace*{0.3cm}
\includegraphics[width=.45\textwidth,angle=0]{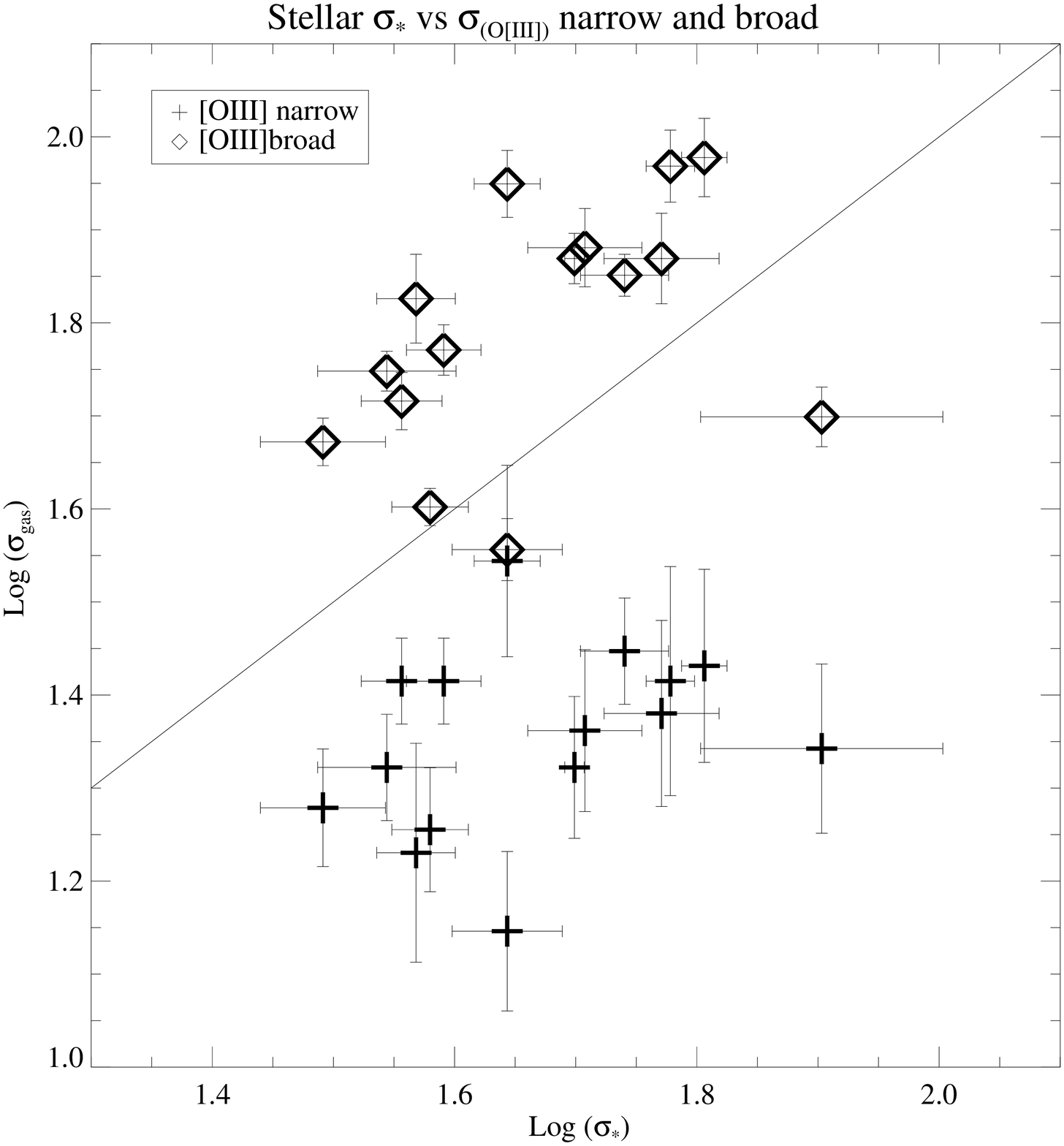}
\caption[]{Gaseous (O[III]) versus stellar velocity dispersions for all the observed circumnuclear regions.  The continuous line represents:  $\sigma _{gas}$ $=$ $\sigma _{stars}$.}
\label{dispersions-comp-c}
\end{figure}

In Fig.\ \ref{dispersions-comp-b} and Fig.\ \ref{dispersions-comp-c} we show the velocity dispersion of the narrow and broad Gaussian components of H$\beta$ and [\OIII], respectively, versus the stellar velocity dispersion. The narrow components of both H$\beta$ and [\OIII] span a similar range in velocity dispersion with an average value of  $\sim$23\,\kms. On the other hand, the values of the velocity dispersion of the broad component of H$\beta$ are similar to those of the stellar velocity dispersion while the broad component of the [\OIII] line shows, in most cases, values of $\sigma$ in excess of the stellar one. The correlation coefficient of the broad component of both H$\beta$ and [\OIII]  with the stellar velocity dispersion is much smaller than 0.5 indicating  basically no correlation.



Bearing in mind that we may over-interpret the results of what is in reality a
small sample, the fact that the narrow component of the gas in the CNSFRs in 3
different galaxies is, inside the error bars, roughly constant  suggests that
it may  be originated in an inner disc that would be mainly supported by
rotation. On the other hand, the stars and the gas responsible for the broad
component could be related to the star-forming regions themselves and
therefore would be mostly supported by dynamical pressure \citep[see][and
references therein]{Pizzella+04}. This could, in principle, correspond to the
gas response to the gravitational potential of the stellar cluster, thus
explaining the coincidence with the stellar velocity dispersion in most cases
for the { broad component of the} H$\beta$ line. To disentangle the origin
of the two different kinematical components it will be necessary to map these
regions with higher spectral and spatial resolution and much better
signal-to-noise ratio in particular for the [\OIII] lines.  

\textbf{
The development of the Integral Field Unit (IFU) instruments to perform 3D
spectroscopy during the last years has provided the spatial
coverage required to study extended galactic or extragalactic star-forming
regions \citep[see
e.g.][]{Monreal-Ibero+10,Monreal-Ibero+11,Rosales-Ortega+10,Rosales-Ortega+11,Garcia-Benito+10,Sanchez+10,Lopez-Sanchez+11,Perez-Montero+11,Kehrig+12,Sanchez+12b,Lopez-Hernandez+13,James+13,Firpo+13}. 
Recently the first results of very powerful surveys of galaxies using IFU
instruments have been released. The Calar Alto Legacy Integral Field Area
(CALIFA) survey is obtaining spatially resolved spectroscopic information of a
selected sample of $\sim$600 galaxies in the Local Universe
\citep{Sanchez+12a}. This survey has an effective spatial resolution of about 
2\,arcsec per fibre, and the final spectral resolution in FWHM is
$\sim$6.5\,\AA, (i.e.\ $\sigma\,\sim\,150\,$\kms) for the V500 grating and
$\sim$2.7\,\AA, (i.e.\ $\sigma\,\sim\,85\,$\kms) for the V1200. These
instrumental configurations are not enough to disentangle the origin of the two
different kinematical components with the required accuracy. The possibility 
of a future survey performed using the Sydney-AAO Multi-object Integral field
spectrograph (SAMI) would be better for our purposes since its higher spectral
resolution is R=$\lambda/\Delta\lambda$\,$\sim$\,13000 (i.e.\
$\sigma\,\sim\,23\,$\kms at 4861\,\AA), and each fibre subtends 1.6\,arcsec on
the sky \citep{Croom+12}. However,
medium or high dispersion  slit spectroscopy are a better option for
spectrophotometry, or when the object is very compact, or even extended but
with few star-forming knots. This is also the case when good spatial and
spectral resolution and  simultaneous wide spectral coverage are required
\citep[see
e.g.][]{Cumming+08,Firpo+10,Firpo+11,Hagele+06,Hagele+07,Hagele+08,HagelePhD,Hagele+09,Hagele+10a,Hagele+11,Hagele+12,Lopez-Sanchez+09,Lopez-Sanchez+10c,Lopez-Sanchez+10b,Lopez-Sanchez+10,Amorin+12}.
}

\subsection{Effects on abundance estimates}

\label{BPT}

The presence of two kinematically segregated gaseous components could have an
effect on the classification of the activity in the central regions of
galaxies through  diagnostic diagrams and also on abundance determinations if
they are made from low dispersion spectra by biasing the  line ratios
involved \textbf{\citep[see e.g.][]{Esteban+92,Lopez-Sanchez+07,Hagele+12,James+13,James+13b}}.  
The only line ratio available from our data for the two kinematically
separated components is [\OIII]/H$\beta$, { whose logarithmic ratios are
  listed in Table \ref{ratio}}. We find that in all the  CNSFRs of 
NGC~2903 and NGC~3351, ([\OIII]/H$\beta$)$_{narrow} <$
([\OIII]/H$\beta$)$_{broad}$ which indicates a higher excitation for the broad
component gas. In the case of NGC~3351 the mean value of log
([\OIII]/H$\beta$)$_{narrow}$ is -1.57 as compared to -0.93 for the same line
ratio of the broad components, while a value of -1.04  is obtained from a
single Gaussian fit (see Fig. 12 in H07). In these  cases the
broad component contributes more than 80\% to the [\OIII] total flux while
this contribution is smaller than 40\% for the H$\beta$ line. This is not the
case for the CNSFRs of NGC~3310 for which the contribution by the broad
component to both the [\OIII] and H$\beta$ emission line fluxes is similar and
about 50\% which maintains the [\OIII]/H$\beta$ ratio almost constant. 


\begin{table}
\centering
{\footnotesize
\caption{Logarithm of the [O{\sc iii}]5007/H$\beta$ line ratio.}
\begin{center}
\begin{tabular} {@{}l c c c@{}}
\hline
\hline
 Region &  \multicolumn{3}{c}{log([O{\sc iii}]5007/H$\beta$)}\\
        &         One component &  Narrow  & Broad   \\
\hline

\multicolumn{4}{c}{NGC\,2903}\\[2pt]

R1+R2  &  -1.03$\pm$0.05   & -1.64$\pm$0.09  & -0.76$\pm$0.10 \\     
R1+R2  &  -0.92$\pm$0.07   & -1.51$\pm$0.11  & -0.57$\pm$0.14 \\     
R4     &  -0.78$\pm$0.11   & -1.01$\pm$0.10  & -0.71$\pm$0.15 \\     
R7     &  -0.78$\pm$0.10   & -1.79$\pm$0.12  & -0.40$\pm$0.18 \\[2pt]

\multicolumn{4}{c}{NGC\,3310}\\[2pt]                                                           
                                                                     	                   
R1+R2   &   0.42$\pm$0.01   &  0.39$\pm$0.01  &  0.45$\pm$0.03  \\      
R4      &   0.28$\pm$0.01   &  0.28$\pm$0.01  &  0.28$\pm$0.04  \\      
R4+R5   &   0.42$\pm$0.01   &  0.39$\pm$0.01  &  0.44$\pm$0.03  \\      
R6      &   0.28$\pm$0.01   &  0.30$\pm$0.01  &  0.28$\pm$0.06  \\      
S6      &   0.21$\pm$0.01   &  0.21$\pm$0.01  &  0.22$\pm$0.08  \\      
R7      &   0.23$\pm$0.01   &  0.21$\pm$0.01  &  0.32$\pm$0.09  \\      
R10     &   0.19$\pm$0.01   &  0.20$\pm$0.04  &  0.20$\pm$0.08  \\[2pt] 

\multicolumn{4}{c}{NGC\,3351}\\[2pt]                                	                   
                                                                     	                   
R2    &  -1.07$\pm$0.06   & -1.66$\pm$0.08  & -0.93$\pm$0.12 \\     
R2    &  -1.01$\pm$0.06   & -1.55$\pm$0.08  & -0.96$\pm$0.13 \\     
R3    &  -1.10$\pm$0.06   & -1.57$\pm$0.07  & -0.93$\pm$0.10 \\     
R3    &  -1.00$\pm$0.06   & -1.52$\pm$0.09  & -0.89$\pm$0.10 \\[2pt]     

\hline
\end{tabular}
\end{center}
\label{ratio}
}
\end{table}

\begin{figure}
\centering
\includegraphics[width=.450\textwidth,angle=0]{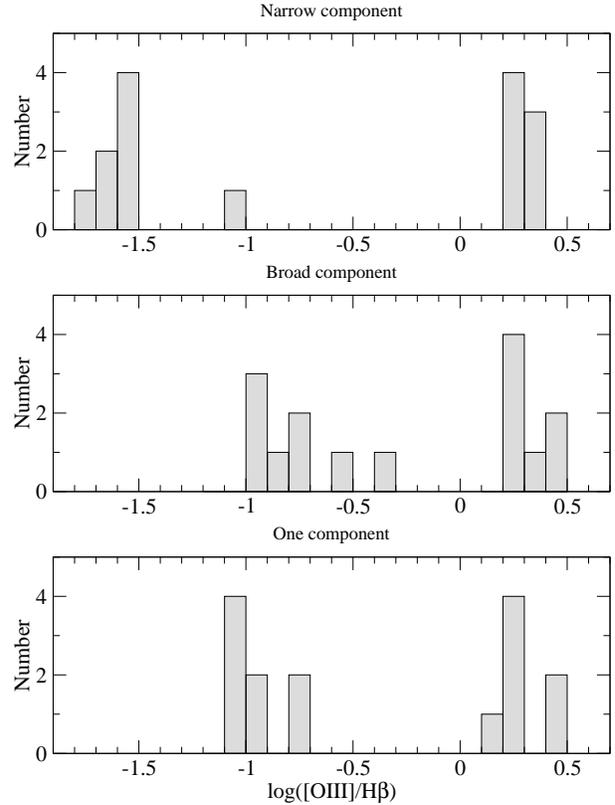}\\\vspace*{0.25cm}
\caption[]{Histogram of the logarithmic [\OIII]/H$\beta$ ratio for the
  different gaseous kinematic components compared to values derived from
  single Gaussian fits.} 
\label{histo}
\end{figure}

Figure \ref{histo} shows a histogram of the logarithmic [\OIII]/H$\beta$ ratio
for the different components of the gas, broad and narrow, and also for the
case of a single Gaussian fit { (one component)}. It can be seen that for
the CNSFRs of NGC\,2903 and NGC\,3351, the gas showing different velocity
dispersions also show different ionization properties with the broad component
one showing a higher degree of excitation { (see also Table \ref{ratio})}. The data with the highest
[\OIII]/H$\beta$ ratio correspond to the regions in NGC\,3310 for which no
differences are found among the different cases.

\begin{figure*}
\centering
\includegraphics[width=.90\textwidth,angle=0]{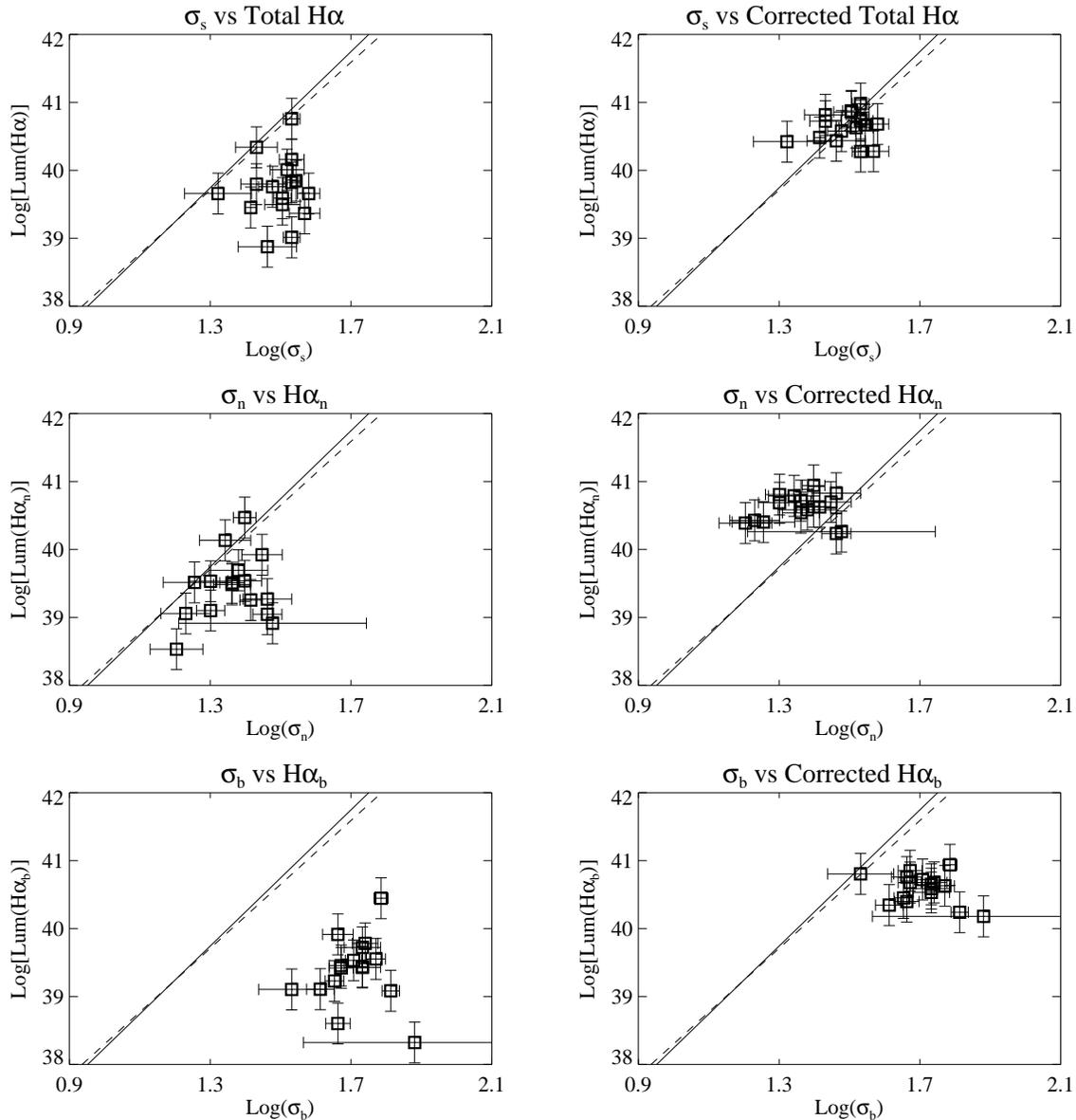}\\\vspace*{0.25cm}
\caption[]{$\log(\sigma) - \log(L)$ relations for the observed CNSFRs.  
  Panels show the relations derived for the velocity dispersions of the gas
 using a single Gaussian fit (top), narrow components (centre) and broad components (bottom), respectively. The panels on the right show the result after correcting the luminosity for evolution  (see text).
  The lines correspond to  a linear fit to ``young'' Giant H{\sc ii} regions,  plotted as a
  reference value; B02 (dashed line) and Ch\'avez et al. 2012 (solid line). }
\label{displum}
\end{figure*}

\subsection{$\sigma$-L and M-L relations}

\label{luminosity}

Fig.\ \ref{displum} shows the distribution of the CNSFRs in the $L({\rm H\alpha}$) vs.\ $\sigma$ plane. 
In the left column, upper panel we plot the logarithm of the total H$\alpha$ luminosity (values taken from the literature corrected for extinction) versus the gas velocity dispersion derived using a single Gaussian fit. In the middle and lower panels we show the same relation but for the H$\alpha$ luminosity of the narrow [L$_n$(H$\alpha$)] and broad [L$_b$(H$\alpha$)] components
of the two-Gaussian fits  derived using the EM$f$ estimated from the fit to
the H$\beta$ emission line, versus the corresponding velocity dispersion given
by the multi-Gaussian fit ($\sigma_n$ and $\sigma_b$, respectively). The
linear fits from Bosch et al.\ (\citeyear{Bosch+02}; hereafter B02) (dashed
line) and Ch\'avez et al (\citeyear{Chavez+12}; 
 hereafter Ch12) (solid line) are also shown. These data correspond to
 ``young'' single clusters in  well studied giant extragalactic \HII\ regions. They comprise the structureless entities in the regions free of bubbles and shells which are supposed to appear at more evolved phases of evolution. These lines may be taken as the locus of virialized structures sampled by these compact emission knots. 
 
Contrary to what is found for the compact emission line knots in nearby
spirals by B02 and Ch12, no correlation is found between $L({\rm H\alpha}$) and $\sigma$ either for single Gaussian fits or for individual kinematical components; only three of our CNSFRs lie on top of the locus of virialized objects  for the single Gaussian fit case. A weak correlation seems to emerge for the narrow kinematical component of hydrogen that is maintained for the broad component although shifted to higher velocity dispersions. At any rate, the large dispersion of the data  seems to be a common feature.
  
A certain degree of scatter in the diagram is expected related to the age evolution of the ionizing regions which would lower their H$\alpha$ luminosity while keeping their velocity dispersion almost constant. The presence of this effect is reinforced by the fact  that the three regions on top of the B02 and Ch12 lines are those with the largest equivalent widths of H$\beta$ (lower age) and therefore should be expected to be the less evolved ones.
  
A correction for this effect can be made by using the same expression
employed for the estimation of the mass of the ionizing clusters, using a
reference value for the equivalent width of H$\beta$ for an age lower than 3
Myr { $EW({\rm H\beta})_0$\,=\,150\,\AA.}
  
\begin{equation}
L({\rm H\alpha})_{\mathrm{Corrected}}\,={L({\rm H\alpha})_{\mathrm{Observed}}}{10^{[44.48\,+\,0.86\,\log\left[EW(H\beta)_0\right]}}
\end{equation}

The  panels on the right column of Fig \ref{displum} show the correlations just discussed once these evolutionary corrections have been applied. The net result is a considerable reduction of the scatter thus suggesting that the equivalent width of H$\beta$ (i.e.~age) is the second parameter underlying the $L({\rm H\alpha}$) vs.\ $\sigma$ relation.

\begin{figure*}
\centering
\includegraphics[width=.55\textwidth,angle=0]{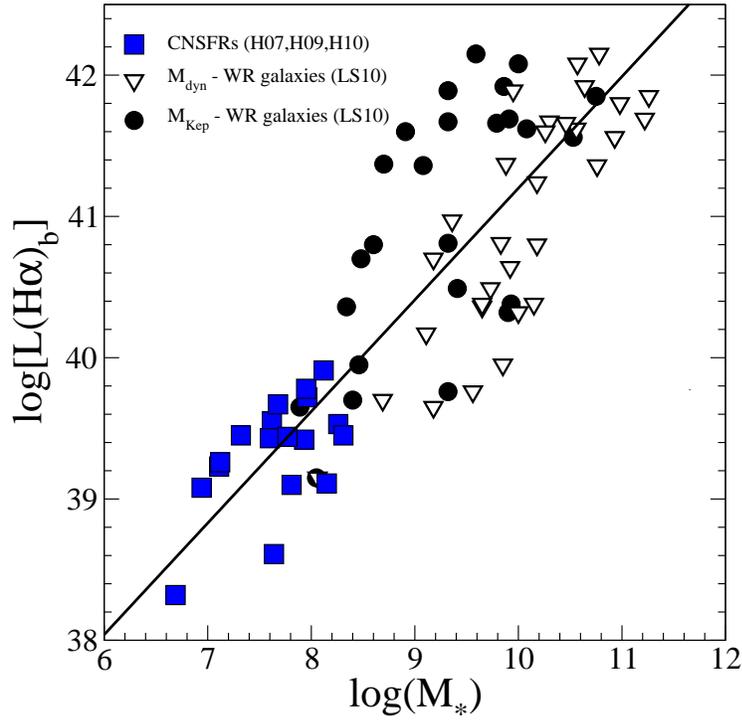}
\caption[]{H$\alpha$ luminosity vs.\ cluster dynamical mass for the observed CNSFRs (squares). Similar data on WR galaxies from the sample of LS10 are shown for comparison. In this latter case both dynamical (inverted triangles) and Keplerian (circles) masses are plotted.}
\label{masslum1}
\end{figure*}

The stellar velocity dispersion behaves similarly to  the broad 
component of the emission lines, albeit with larger dispersion.

\begin{figure*}
\centering
\includegraphics[width=.55\textwidth,angle=0]{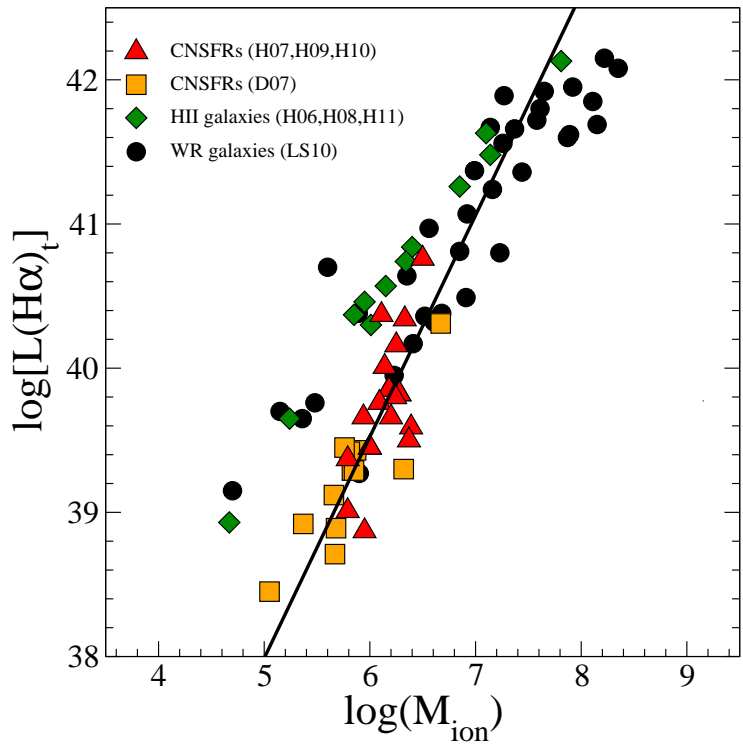}\hspace*{0.15cm}
\caption[]{H$\alpha$ luminosity vs.~ionizing star cluster mass  for the
  observed CNSFRs (triangles and squares). Similar data on WR galaxies (solid
  circles) and \HII galaxies (diamonds) are also shown. }
\label{masslum2}
\end{figure*}

\begin{figure*}
\centering
\includegraphics[width=.55\textwidth,angle=0]{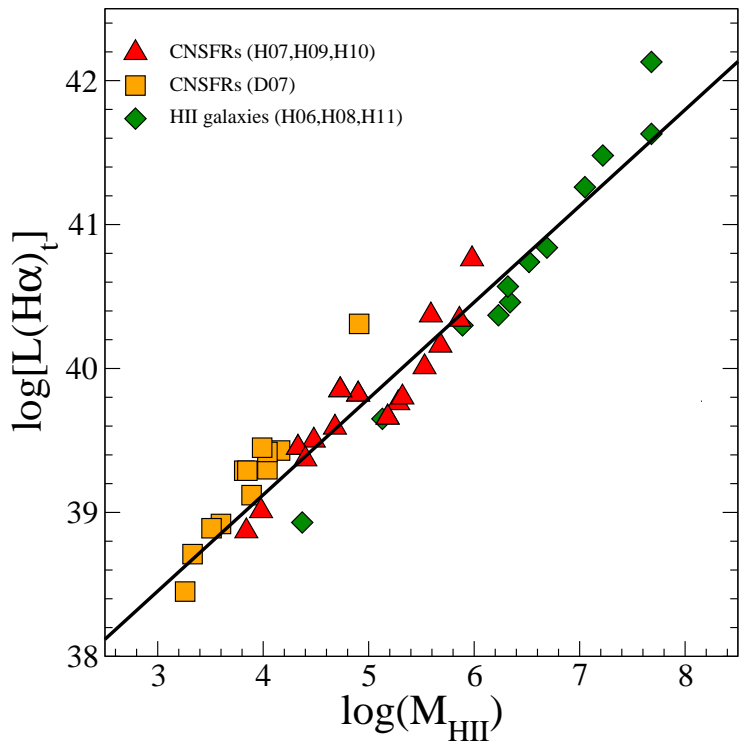}
\caption[]{H$\alpha$ luminosity vs.\ ionized hydrogen mass mass  for the
  observed CNSFRs (squares and triangles). Similar data on \HII galaxies
  (diamonds) are also shown.} 
\label{masslum3}
\end{figure*}

It  is of major interest to find out how widespread is the presence of two distinct components in the emission lines of these ionised regions.
Firstly, a change in position in the diagnostic diagrams  would certainly affect the classification of the activity in the central regions of the concerned galaxies. Secondly,  it will affect the inferences about the nature of the source of ionisation in the two components. Thirdly, it could have an influence on the gas abundance determinations given  that the ratio of the narrow to the broad components of H$\beta$ is, in most
cases, about 1/2 . 

Clearly it is not possible to use global line ratios to estimate gaseous abundances if the permitted and forbidden line fluxes are originated in different kinematic systems.

\label{masses}

We analyse the relation between the H$\alpha$ luminosities of the broad
component and the
derived masses for the CNSFRs.  Fig.\ \ref{masslum1} shows the
relation between the logarithm of the upper limit to the dynamical masses and
$\log(L({\rm H\alpha})_b)$, together with the dynamical ($M_{dyn}$) and Keplerian
($M_{Kep}$) masses of Wolf-Rayet (WR) galaxies derived by \cite[][hereafter
LS10]{Lopez-Sanchez+10}. Values of $M_{dyn}$ were estimated from \HI radio
observations considering the inclination-corrected maximum rotation velocity
which is obtained at the maximum radius observed in their deep optical images,
and assuming  virial equilibrium. As was pointed out by LS10, since the
extension of the neutral gas is usually larger than the extension of the
stellar component, their values of $M_{dyn}$ may be underestimated, although
they can be used as a rough estimation of the total
mass of the galaxies. The $M_{Kep}$ were estimated from the kinematics
of the ionised gas. 
The Keplerian masses are lower than the dynamical ones, as expected, in almost all
cases (see Fig.\ 11 of LS10), with the more massive galaxies showing 
higher $M_{Kep}$/$M_{dyn}$ ratios. This fact indicates that the kinematics
derived from the ionised gas is not completely appropriate for deriving the total
dynamical mass in this kind of objects (LS10). 

The upper limits to the dynamical masses of the CNSF complexes seem to follow
a sequence when we compare them with those masses derived for the WR
galaxies (see  Fig.\ \ref{masslum1}). Our objects are located in
the lower part of the sequence, showing lower masses and lower H$\alpha$
luminosities. 
A fitting to the logarithmic dynamical mass-luminosity
(M-L) relation shown by the CNSFRs 
(solid line in the figure) gives the following
expression:  

\begin{equation}
\log(L({\rm H\alpha})_b)\,=\,(0.8\,\pm\,0.1)\,\log(M_{\ast})\,+\,(33\,\pm\,1).
\end{equation}


%

%
%
%

The relation between the masses of the ionising clusters of the CNSFRs and
their H$\alpha$ luminosities is shown in 
Fig.\ \ref{masslum2}. In spite of depending on a theoretical relationship which
is only function of the H$\alpha$ luminosity and the equivalent width of the
H$\beta$ emission line we can appreciate that the CNSFRs (including the
objects studied by D07) seem to define a different  sequence to the one followed by the
\HII\ galaxies studied in \citeauthor{Hagele+06}
(\citeyear{Hagele+06,Hagele+08,Hagele+11}, hereafter H06, H08 and H11), and the
WR galaxies analysed by 
LS10 \cite[see][for a detailed description of the optical
data]{Lopez-Sanchez+08}. This could be due to underestimating the
EW(H$\beta$) hence overestimating the masses of the
ionising star clusters. The dilution of the EW(H$\beta$) can be caused by the 
presence of an old underlying stellar population with wider H$\beta$
absorption that rises the continuum and depresses the emission line {(see a
detailed discussion about this effect in D07)}. 
In addition, the low EW(H$\beta$) values for CNSFRs might
come from the contribution to the observed continuum by the galactic bulge
population. \cite{Dors+08} suggest that the combination of an underlying older
population from the region itself and stars from the bulge could account for
the observed effect. The strong influence that the measured equivalent width
has on the derived mass of the ionising cluster indicates that the trend
observed in Figs.\ \ref{masslum1}, \ref{masslum2} and \ref{masslum3} can be attributed to the fact that the
contamination by the bulge will be relatively stronger as the
intrinsic luminosity of the starburst diminishes. 

This effect is more evident for the less massive
objects. The most massive CNSFRs tend to be located around the position
occupied by the \HII and WR galaxies of similar cluster ionising masses. 
The effect of the presence of an underlying old stellar population over the
equivalent width of H$\beta$ in this kind of emission line galaxies with
recent episodes of star formations is relatively small since the continuum of
the bright young stellar population dominates the light in this spectral range
[see a detailed discussion about the contributions of the different stellar
populations to the EW(H$\beta$) in \citealt{Perez-Montero+10} and H11].

The solid line in  Fig.\ \ref{masslum2} represents the fit to
the logarithmic relation between $M_{ion}$ and total H$\alpha$ luminosities
for the CNSFRs, and is given by
\begin{equation}
\log(L({\rm H\alpha})_t)\,=\,(1.5\,\pm\,0.4)\,\log(M_{ion}) + (30\,\pm\,2).
\end{equation}

%



Finally, in  Fig.\ \ref{masslum3} we have plotted the
relation between the masses of the ionised gas and the total H$\alpha$
luminosities, together with the CNSFRs presented in D07 and the \HII galaxies
studied in H06, H08 and H11. 
The fitting to this relation for the CNSFRs gives
\begin{equation}
\log(L({\rm H\alpha})_t)\,=\,(0.67\,\pm\,0.07)\,\log(M_{{\rm HII}}) + (36.4\,\pm\,0.3)
\end{equation}

%


\section{Summary and conclusions}
\label{summary}

We have analysed the results derived by H07, H09 and H10 using high
spectral resolution data of 17 CNSFRs belonging to three barred
spiral galaxies: NGC\,2903, NGC\,3310 and NGC\,3351. The stellar velocity
dispersions were measured from the CaT lines at $\lambda\lambda$\,8494, 8542,
8662\,\AA. The gas velocity dispersions have been measured
by Gaussian fits to the H$\beta$\,$\lambda$\,4861\,\AA\ and the [O{\sc
    iii}]\,$\lambda$\,5007\AA\ emission lines. 

Stellar velocity dispersions range between 31 and 80\,km\,s$^{-1}$.
In the case of NGC\,2903 and NGC\,3351 ($\sigma_\ast$ between
37 and 64, and between 39 and 66\,km\,s$^{-1}$, respectively) these values are
about 25\,km\,s$^{-1}$ larger than those measured for the gas from the
H$\beta$ emission line using a single Gaussian fit. For NGC\,3310 the stellar
(between 31 and 80\,km\,s$^{-1}$) and gas (H$\beta$) velocity dispersions are in
relatively good agreement, the  stellar ones being marginally larger. 

Single Gaussian fitting to the [O{\sc iii}]\,5007\,\AA\ line provides gas velocity dispersions for NGC\,2903 and NGC\,3351 that are very similar to - or marginally larger than - the stellar ones. The gas velocity dispersion derived from a single Gaussian fit for the [O{\sc iii}] line in
NGC\,3310  is also comparable to the stellar velocity dispersion but, contrary to what was found in the other two galaxies, its value is very similar also to that derived using H$\beta$.

The best Gaussian fits for H$\beta$, however, involved two different components for the gas:
a ``broad component'' with a velocity dispersion similar to that measured for
the stars for NGC\,2903 and NGC\,3351, and about 20\,km\,s$^{-1}$ larger than the stellar one 
for NGC\,3310, and a ``narrow component'' with a velocity dispersion lower than
the stellar one by about 30\,km\,s$^{-1}$. This narrow component shows a
relatively constant value for the two  emission lines (H$\beta$ and
[\OIII]), close to 23\,km\,s$^{-1}$ (with scatter errors of 1.1 and 1.4\,\kms,
respectively) for all the studied CNSFRs. 

The [O{\sc iii}]/H$\beta$ ratio distribution shows that
the two systems are clearly segregated for the high-metallicity
regions of NGC\,2903 and NGC\,3351, with the narrow component
having the lowest excitation. In the regions of the
low-metallicity galaxy, NGC\,3310, these two components and those values
derived using the single Gaussian fit are very similar.

Values for the upper limits to the dynamical masses estimated from the stellar
velocity dispersion using the virial theorem for the studied CNSF complexes
are in the range between 4.9\,$\times$\,10$^6$ and 1.9\,$\times$\,10$^8$\,$M_\odot$.
Masses of the ionising stellar clusters of the CNSFRs have been derived from
their H$\alpha$ luminosities and the equivalent width of the H$\beta$ emission
line under the assumption that the regions are
ionisation bound and have a single stellar population, and without taking into
account any photon absorption by dust. These masses of the regions studied in
the three galaxies vary between 8.0\,$\times$\,10$^5$ and 
4.9\,$\times\,10^6$\,$M_\odot$. Therefore, the ratio of the ionising
stellar population to the total dynamical mass, under these hypothesis, is
between 0.01 and 0.16. The derived masses for the ionised gas, also from their
H$\alpha$ luminosities, vary between 7.0\,$\times$\,10$^3$ and
7.2\,$\times$\,10$^5$\,$M_\odot$.

The distribution of the CNSFRs in the $\log(\sigma) - \log(L)$ plane presents a
correlation between their luminosities and velocity dispersions, albeit
with large dispersion. The single Gaussian fit for our CNSFRs are
slightly shifted to lower luminosities and/or higher values of the gas velocity
dispersion with respect to the regression found for virialised systems,
although the general behaviour is similar to what was found by
\cite{Firpo+10,Firpo+11} for giant extragalactic \HII\ regions of NGC\,6070
and NGC\,7479, and star-forming regions belonging to the Haro\,15 galaxy,
respectively. The narrow components of the multi-Gaussian fits to the emission
lines of the CNSFRs lie very close to this linear regression, with the brightest
objects around the regression, while the broad components are located in a
parallel sequence shifted towards lower luminosities and/or higher velocity
dispersions. We have been able to considerably reduce the observed scatter in
the L-$\sigma$ relation applying an evolutionary correction that depends on
the age of the regions through their equivalent widths of H$\beta$. The
EW(H$\beta$) seems to be a second order parameter underlying this relation.
Although the stellar velocity dispersions have a large
scatter, they present a similar behaviour to the broad component ones. 

We have also analysed the relations between the total H$\alpha$ luminosities
and the derived masses for the studied CNSFRs. The upper limits to the
dynamical masses of the CNSF complexes seem to be located in
the lower part of a sequence showing lower masses and lower H$\alpha$
luminosities than those masses derived
for WR galaxies by LS10. The distribution of the masses of the ionising clusters of the
CNSFRs, including the regions presented by D07, falls outside a
sequence defined by the \HII and WR galaxies analysed by H06, H08 and H11, and
by LS10, respectively. The CNSFRs are also arranged in a steeper slope
sequence which is located toward lower values of the H$\alpha$ luminosities or 
higher masses, except for the most massive regions. This occurs although the
ionising stellar masses depend on a theoretical relationship which is a
function only of the H$\alpha$ luminosity and the equivalent width of the H$\beta$
emission line. Underestimating
EW(H$\beta$) would produce an overestimation of these masses. We have to
remark that, according to our findings, the super star-clusters in CNSFRs seem to contain
composite stellar populations (see e.g.\ H07 and D07). 
The contribution of stars from the bulge projected along the line of
sight \citep{Dors+08} has also to be considered. Although the youngest population of star-forming
complexes dominates the UV light and is responsible for the gas ionisation, it
represents only about 10 per cent of the total mass belonging to the
region. This can explain the low EWs of the emission lines generally measured in these
regions.  This may well apply to the case of other super star-clusters and therefore
conclusions drawn from fits of single stellar population models should
be taken with caution \citep[e.g.\ ][]{McCrady+03,Larsen+04}. On the other
hand, the relation between the derived masses of the ionised gas of the CNSFRs
and the H$\alpha$ luminosities follow
the same sequence defined by the \HII galaxies presented by H06, H08 and H11.

A possible  scenario for understanding  the behaviour of CNSFRs in the
L-$\sigma$ and $\sigma_{gas}$-$\sigma_*$ diagrams is to assume that the narrow
component of hydrogen belongs to an inner rotating disc, therefore its
velocity dispersion is approximately constant for all the CNSFRs. On the other
hand, the broad component of hydrogen would be related to the CNSF complexes
themselves, and therefore its velocity dispersion would be expected to be
similar to the stellar one. (The velocity dispersion of collisionally excited
[\OIII] is in some cases larger than this, a behaviour also seen in Seyfert 2
galaxies, but not in Seyfert 1 \cite[see][]{Jimenez-Benito+00}. In that case,
all the H$\alpha$ derived quantities for the CNSFRs should use the luminosity
of the broad component which is about one half of the total.  
The fact that there is a distinctive behaviour in the $L({\rm H\alpha}$) vs.\ 
$M_{ion}$ in disk \HII regions and CNSFRs in the sense of showing a larger
mass for a given H$\alpha$ luminosity could be related to the \HII regions
being matter bounded, that is photons are escaping the region. Incidentally
(or not so) {these} photons would ionise the hydrogen in the inner disc
and therefore would {in part} be responsible for the narrow component of hydrogen. All
this would be consistent with the picture of some CNSFRs undergoing
``residual'' star formation that involves about 10\% of the total dynamical
mass. This behaviour would be related to the evolution of the region with the
ones showing the larger EW(H$\beta$), that is the younger ones, having more
hydrogen available and thus being less prone to photon escape. The SFRs for
the CNSFRs should also be derived from the luminosity of the broad component
of H$\alpha$ and would be about half the ones derived from the total
H$\alpha$ fluxes. This would explain the lack of detection at radio wavelengths
\citep{Hagele+10b}. 


The existence of more than one velocity component in the ionised gas
corresponding to kinematically distinct systems, deserves further
study. Several results derived from the observations of the different emission
lines could be affected, among others: the classification of the activity in
the central regions of galaxies, the inferences about the nature of the source
of ionisation, the gas abundance determinations, the number of ionising
photons from a given region and any quantity derived from them, as well as the
$\sigma$-L and M-L relations. To disentangle the origin of these two
components and to discriminate among the different stellar population
contributions to the continuum of the regions it will be necessary to map
these regions with high spectral and spatial resolution and much better S/N
ratio. High resolution {2D} spectroscopy with IFUs would be the ideal tool to
approach this issue.

\section*{Acknowledgements}
\label{Acknoledgement}
{We are grateful to the referee, \'Angel L\'opez-S\'anchez, for a thorough
  reading of the manuscript and for his constructive comments and suggestions.}

The WHT is operated in the island of La Palma by the Isaac Newton Group
in the Spanish Observatorio del Roque de los Muchachos of the Instituto
de Astrof\'\i sica de Canarias. We thank the Spanish allocation committee
(CAT) for awarding observing time.

Some of the data presented in this paper were obtained from the
Multimission Archive at the Space Telescope Science Institute (MAST). STScI is
operated by the Association of Universities for Research in Astronomy, Inc.,
under NASA contract NAS5-26555. Support for MAST for non-HST data is provided
by the NASA Office of Space Science via grant NAG5-7584 and by other grants
and contracts.

This research has made use of the NASA/IPAC Extragalactic Database (NED) which
is operated by the Jet Propulsion Laboratory, California Institute of
Technology, under contract with the National Aeronautics and Space
Administration and of the SIMBAD database, operated at CDS,
Strasbourg, France. 

Financial support for this work has been provided by the Spanish
\emph{Ministerio de Educaci\'on y Ciencia} (AYA2007-67965-C03-03 and
AYA2010-21887-C04-03). Partial  support from the Comunidad de Madrid under
grant S2009/ESP-1496 (ASTROMADRID) is acknowledged. ET and RT are grateful to
the Mexican Research Council (CONACYT) for support under grants
CB2005-01-49847F and CB2008-01-103365.

\bibliographystyle{mn2e}
\bibliography{references}

\end{document}